\documentstyle[12pt]{article}

\def\fff{\vrule width0.5pt height5pt depth1pt}
\def\pp{{{ =\hskip-3.75pt{\fff}}\hskip3.75pt }}

\def\mach{\vartheta}

\begin{document}

\begin{titlepage}

\begin{flushright}
QMW-PH-97-25\\
hep-th/9708047
\end{flushright}

\vspace{3cm}

\begin{center}

{\bf \Large The Gauged (2,1) Heterotic Sigma-Model}

\vspace{.7cm}

Mohab Abou Zeid and Christopher M.\ Hull

\vspace{.7cm}

{\em Physics Department, Queen Mary and Westfield College, \\
Mile End Road, London E1 4NS, U.\ K.\ }

\vspace{2cm}

\vspace{1cm} August 1997

\vspace{1cm}

\begin{abstract}

The geometry of (2,1) supersymmetric sigma-models with
isometry symmetries is discussed. The gauging of such symmetries
in superspace is then studied. We find that the coupling to the
(2,1) Yang-Mills supermultiplet can be achieved provided certain geometric
conditions are satisfied. We construct the general gauged action, using an
auxiliary vector to
generate  the full
non-polynomial structure.

\end{abstract}

\end{center}

\end{titlepage}

\section{Introduction}                          \label{intro}

        It has  been proposed by Kutasov and Martinec~\cite{KutMart}
that the heterotic strings with (2,1) world-sheet supersymmetry provide an
appropriate framework for implementing and extending an earlier idea due to
Green~\cite{Mike}, who suggested that the world-sheet theories of various
string theories are obtainable from the target space theories
of two-dimensional strings. Furthermore, the authors of~\cite{KutMart}
showed that different
target vacua for (2,1) heterotic strings correspond to the type IIB string, to
the membrane of M theory and to their compactifications.

        The $N=2$ strings, and especially the (2,1) heterotic strings, have
a number of remarkable features~\cite{OogVaf} whose consideration within the
framework of string and M theory duality lead to the above picture. First,
their spectrum is simple in that it contains only a finite number of massless
modes (there is no tower of massive modes as in other critical string
theories),
and their interactions are likewise simple, as all $n$-point scattering
amplitudes
vanish for $n\geq 4$. Although their
critical
dimension is four, supersymmetry implies that the signature of the target
manifold is either $4+0$ or $2+2$, the latter case being relevant to the
heterotic
theory. Furthermore, the $N=2$ superconformal algebra contains a $U(1)$ current
whose left-moving component must be gauged in the (2,1) heterotic version of
the theory~\cite{OogVaf}. This in turn necessitates the introduction of a new
set of ghosts which raises the critical dimension in the left-moving sector by
2; hence the theory, when embedded in ten-dimensional spacetime, contains a
left-moving internal $N=1$ SCFT with $\hat{c} =8$. In order to
ensure the absence of BRST anomalies one must require the left-moving $U(1)$
current to be of the form $v \cdot \partial X$ where $v$ is a null
Killing vector, $v^2 =0$. If the component of this vector along the internal
directions
vanishes, the string theory lives effectively in $1+1$ dimensions, and one can
recover the bosonic, type II and heterotic world-sheet theories in
a physical gauge for different choices of the internal $\hat{c}=8$
SCFT~\cite{KutMart}.
On the other hand, if the vector $v$ lies partly in the internal sector, one
obtains an effective $2+1$ theory that corresponds to the world-volume theory
of the supermembrane of eleven-dimensional supegravity~\cite{KutMart}. It was
further suggested that a similar construction could yield other $p$-branes and
that the (2,1) heterotic string is the unifying structure that underlies all
M theory vacua~\cite{KutMart}. Also, consideration of the scattering amplitudes
of the (2,1) heterotic string has led to the construction of the exact
classical
action for the target field theory (see refs.~\cite{KM2,CHlat}). This
describes the dynamics of a set of self-dual Yang-Mills fields coupled to
self-dual gravity~\cite{OogVaf}.

        In this context, it is important to study the geometry of the gauged
two-dimensional (2,1) supersymmetric sigma-models, which describe the
propagation
of (2,1) heterotic strings on certain hermitian manifolds (with torsion)
that admit isometry symmetries. The information
obtained concerning the geometry and quantum dynamics of the sigma-model
should
 be useful in further studies of the target field theory of the (2,1)
heterotic string.

        The general two-dimensional non-linear sigma-model with two
right-moving and one left-moving supersymmetries was considered
in~\cite{Hull21}
(following ref.~\cite{HullWitt}) and was formulated in a
superspace with (1,1) supersymmetry manifest. The geometric conditions the
model must satisfy in order to
possess a second right-moving supersymmetry and to be finite at one loop were
found~\cite{Hull21}. A formulation of the model in (2,1) extended superspace
was given in~\cite{DinSeib}, while an alternative extended superspace
formulation
was proposed by Howe and Papadopouolos~\cite{HP}. In this \lq HP-formalism',
the action  is an integral with the
usual (1,1) superspace measure $\int d^2 \sigma d^2 \theta$ of a Lagrangian
constructed from (2,1) superfields. Checking that the action is
independent of the extra $\theta$ is necessary to show
(2,1) supersymmetry.

The gauging of  isometry
symmetries of the (2,1) sigma-model in the HP formalism was achieved in
ref.~\cite{gauging}.
  However, only (1,1) supersymmetry
was manifest in this approach and  for many purposes, such as the coupling to
supergravity, an approach based on a conventional superspace formalism is more
convenient. The purpose of this paper is to construct an alternative
formulation of the gauged (2,1) sigma-model with torsion.
The problem is that the action has a complicated non-polynomial dependence on
the
scalar gauge prepotentials which is hard to find directly.
There is a similar non-polynomial structure in the $N=1$ sigma-model in four
dimensions and in the
(2,2) sigma-model in two dimensions.
There the ungauged Lagrangian is given in terms of a K\"{a}hler potential and
the
isometry in general
changes this by a K\"{a}hler gauge transformation, $K(z,\bar z) \to K(z,\bar z)
+f(z)+\bar f(\bar z)$.
In
ref.~\cite{in&out},  an extra    superfield coordinate was introduced to
construct a higher-dimensional target space in which the isometry became a
conventional symmetry of
the Lagrangian that could be gauged by minimal coupling. Eliminating the extra
  superfield coordinate
then generated the non-polynomial structure of the gauged action.
This method was later used in
ref.~\cite{gauging} to construct the gauged (2,2) sigma-model with torsion in
two dimensions, and
our purpose here is to develop it further to derive the full non-polynomial
structure of the gauged
(2,1) supersymmetric sigma-model.
This generalises the results of
 our previous paper~\cite{AH1} in which
the gauging was achieved  for a special class of (2,1) target space geometries
with isometries.

        The paper is organised as follows. In section~\ref{review} we review
the
(1,1) supersymmetric non-linear sigma-model with Wess-Zumino term and the
geometric conditions for the model to be invariant under an extra chiral
supersymmetry. This introduces some notation and conventions which we use to
formulate the (2,1) supersymmetric model in (2,1) superspace,
following~\cite{HullWitt,Hull21,DinSeib}. We then review the analysis of the
conditions under which the (1,1) and the (2,1) supersymmetric models have
isometry symmetries and introduce certain potentials
which will play a central role in the gauging of these models in
section~\ref{kinematics}. A more detailed discussion was recently given
in~\cite{AH1}, and we refer the
reader to that paper for details of the construction. We begin our
discussion of the gauging in section~\ref{11gauging} by recalling some of the
results of~\cite{gauging}, where the (1,1) and (2,1) supersymmetric models
were coupled to the (1,1) and (2,1) Yang-Mills supermultiplets using the
approach of~\cite{HP}. In this approach, only (1,1) supersymmetry
is manifest. As
explained above, this has a number of disadvantages, which leads us to seek
a new form of the gauged action for the (2,1) heterotic model with manifest
(2,1) supersymmetry. In section~\ref{SYMetc}
we present the (2,1) Yang-Mills supermultiplet
and   discuss  the transformation properties of the scalar and gauge multiplets
under the gauge and the isometry groups.  In sections~\ref{notheta},
\ref{Noether} and~\ref{trick},
we give an alternative formulation of the gauged (2,1) model, utilizing and
adapting
the approach of~\cite{in&out}. The gauged superspace action is constructed
first
for a special class of models in section~\ref{notheta}. The generic models are
then considered in sections~\ref{Noether} and~\ref{trick}, and a new gauged
superspace action is
found. We conclude in section~\ref{concl}, where we briefly mention some
possible applications of our work in the context of the recent
developments referred to above.

\section{The (2,1) Heterotic Sigma-Model}  \label{review}

        Consider first the general non-linear sigma-model with (1,1)
supersymmetry   with
 a Wess-Zumino term~\cite{HullWitt,C&B}. The (1,1)
superspace is parametrised by two Bose coordinates $\sigma^{\pp} , 
\sigma^{=}$ and two Fermi
coordinates $\theta_{1\pm}$ of opposite chirality. The (1,1) superspace action
for
this model is~\cite{GHR}
\begin{equation}
S_{(1,1)}=\frac{1}{4i}\int d^2 \sigma d^2 \theta \left[ g_{ij}(\phi ) +b_{ij}
(\phi ) \right] D_{1+}\phi^{i} D_{1-}\phi^{j}  ,
\label{S11}
\end{equation}
where the $\phi^i$, $i=1\ldots d$, can be viewed as coordinates on some
$d$-dimensional manifold $M$ with metric $g_{ij}$ and torsion $H$ given by the
curl of the antisymmetric tensor $b_{ij}$,
\begin{equation}
H_{ijk} =\frac{1}{2}\left( b_{ij,k}+b_{jk,i}+b_{ki,j}\right)  ;
\label{torsion}
\end{equation}
here
\begin{equation}
D_{1+}=\frac{\partial}{\partial \theta_{1+}} +i
\theta_{1+}\frac{\partial}{\partial \sigma^{\pp}}  \ \ ,\ \ 
D_{1-}=\frac{\partial}{\partial \theta_{1-}} +i
\theta_{1-}\frac{\partial}{\partial \sigma^{=}}.
\label{superD1}
\end{equation}
The action~(\ref{S11}) is manifestly invariant under (1,1) supersymmetry,
target space general coordinate transformations and antisymmetric tensor
gauge transformations $\delta b_{ij}
=\partial_{[i}\lambda_{j]}$. Furthermore, it was shown
in~\cite{HullWitt,GHR,in&out}
that~(\ref{S11}) will be invariant under an extra chiral supersymmetry
\begin{equation}
\delta \phi^i = J^i{}_j (\phi ) \varepsilon_-   D_{1+}\phi^j
\label{extraSUSY}
\end{equation}
and so have (2,1) supersymmetry provided that $d$ is even and that (i)
$J^i{}_j$
is a complex structure satisfying
\begin{eqnarray}
J^i{}_{j} J^j{}_{k} & = & -\delta^{i}_{k} \nonumber \\ N^k{}_{ij} & \equiv &
J^l{}_{i}J^k{}_{[j,l]} -J^l{}_{j}J^k{}_{[i,l]} = 0 \label{Jcstruct}
\end{eqnarray}
(ii) $J^i{}_{j}$ is covariantly constant
\begin{equation}
\nabla_i J^j{}_k \equiv J^j{}_{k,i} +\Gamma^{j}_{il}J^l{}_{k}
-\Gamma^{l}_{ik}J^j{}_{l} = 0
\label{Jcovconst}
\end{equation}
with respect to the connection
\begin{equation}
\Gamma^{i}_{jk} = \left\{ \begin{array}{c} i \\ jk \end{array} \right\}
+g^{il}H_{jkl}
\end{equation}
which differs from the usual Christoffel connection by the (gauge invariant)
totally antisymmetric torsion~(\ref{torsion})
and (iii) the metric $g_{ij}$ is hermitian with respect to the complex
structure,
\begin{equation}
g_{ij}J^i{}_{k} J^j{}_{l} =g_{kl}  .         \label{ghermJ}
\end{equation}

        In a complex coordinate system $z^\alpha$,
$\overline{z}^{\overline{\beta}}=(z^\beta )^*$, $(\alpha ,
\overline{\beta}=1\ldots \frac{1}{2}d)$ in which the line element is $ds^2
=2g_{\alpha \overline{\beta}}dz^{\alpha} d \overline{z}^{\overline{\beta}}$
and the complex structure is constant and diagonal,
\begin{equation}
J^i{}_j  =  i\left( \begin{array}{cc} \delta^{\beta}{}_\alpha & 0  \\
0 & -\delta^{\overline{\beta}}{}_{\overline{\alpha}} \end{array} \right)  ,
\end{equation}
these conditions imply that the torsion is given by
\begin{equation}
H_{\alpha \beta \overline{\gamma}}=\frac{1}{2}\left( g_{\alpha
\overline{\gamma},
\beta}-g_{\beta \overline{\gamma},\alpha}\right) \ \ , \ \ H_{\alpha \beta
\gamma}=0
\end{equation}
and that the metric satisfies
\begin{equation}
g_{\alpha [ \overline{\beta},\overline{\gamma}]\delta} -g_{\delta [
\overline{\beta},\overline{\gamma}]\alpha} = 0  .
\end{equation}
Then the geometry is determined locally by some vector field $k_\alpha
(z,\overline{z})$,
\begin{eqnarray}
g_{\alpha \overline{\beta}} & = & \partial_{\alpha}
\overline{k}_{\overline{\beta}}+
\partial_{\overline{\beta}}k_{\alpha} \nonumber \\ b_{\alpha \overline{\beta}}
& = &
\partial_{\alpha}\overline{k}_{\overline{\beta}}-\partial_{\overline{\beta}}
k_{\alpha} \nonumber \\ H_{\alpha \beta \overline{\gamma}} & = & \frac{1}{2}
\partial_{\overline{\gamma}} \left( \partial_{\alpha}k_{\beta}
-\partial_{\beta}
k_{\alpha}\right)  .
\label{geometry}
\end{eqnarray}
If the torsion $H=0$, the manifold $M$ is K\"{a}hler with $k_\alpha =
\frac{\partial}{\partial z^{\alpha}}K(z ,\overline{z})$
where $K(z ,\overline{z})$ is the K\"{a}hler potential, and the
(2,1) supersymmetric model in fact has (2,2) supersymmetry, while for $H
\pp 0$, $M$ is a hermitian manifold with torsion of the type
introduced in~\cite{HullWitt,GHR}.

        We now wish to formulate the (2,1) model directly in the (2,1)
superspace. The latter is parametrised by $\sigma^\pp ,\sigma^{=}$, two real Fermi
coordinates
of the same chirality $\theta_{1+}$ and $\theta_{2+}$, and a single real Fermi
coordinate $\theta_{-}$ of the opposite chirality. $\theta_{1+}$ and
$\theta_{2+}$
can be combined into a complex coordinate $\theta_{+}=(\theta_{1+}+
i\theta_{2+})/\sqrt{2}$ and its complex conjugate
$\overline{\theta}_{+}=(\theta_{1+}
-i\theta_{2+})/\sqrt{2}$, and it is natural to define the complex conjugate
supercovariant derivatives
\begin{equation}
D_+ = \frac{1}{\sqrt{2}}\left( D_{1+}+iD_{2+}\right) \ \ , \ \ \overline{D}_+
= \frac{1}{\sqrt{2}}\left( D_{1+}-iD_{2+}\right)
\end{equation}
with $D_{1+}$ and $D_{2+}$ as in~(\ref{superD1}). The supersymmetric sigma
model can then be formulated in (2,1) superspace in terms of scalar
superfields\footnote{We will use the notation $\varphi$ for (2,1) scalar
superfields to distinguish them from the (1,1) scalar superfields $\phi$.}
$\varphi^\alpha$ which are constrained to satisfy the chirality conditions
\begin{equation}
\overline{D}_{+}\varphi^\alpha =0 \ \ , \ \ D_+
\overline{\varphi}^{\overline{\alpha}} =0  .         \label{chiral}
\end{equation}
The lowest components of the superfields
$\varphi^\alpha |_{\theta =0} =z^\alpha$ are the bosonic complex coordinates
of the space-time. The most general renormalizable and Lorentz invariant (2,1)
superspace action is then~\cite{DinSeib}
\begin{equation}
S =i\int d^2 \sigma d\theta_+ d\overline{\theta}_+ d\theta_- \left( k_\alpha
D_- \varphi^\alpha -\overline{k}_{\overline{\alpha}}D_-
\overline{\varphi}^{\overline{\alpha}} \right)  ,
\label{21action}
\end{equation}
and will be gauged in sections~\ref{notheta}, \ref{Noether} and~\ref{trick}. At
this stage, we
simply note that the relations~(\ref{geometry}) which determine the geometry in
terms of the vector field $k_\alpha$ can be recovered from~(\ref{21action}) by
performing the integration over $\theta_{2+}$ in the usual way and identifying
the resulting (1,1) superspace action with~(\ref{S11}), where $g_{ij}$
and $b_{ij}$ are given by eq.~(\ref{geometry}). Also, notice that
the geometry is left invariant by the following transformation
\begin{equation}
\delta k_\alpha = \rho_\alpha  \label{symm}
\end{equation}
provided $\rho_\alpha$ satisfies
\begin{equation}
\overline{\partial}_{\overline{\beta}}\rho_\alpha =
i\partial_\alpha \overline{\partial}_{\overline{\beta}}\chi
 \end{equation}
for some arbitrary real $\chi$.
This implies that $\rho$ is of the form
\begin{equation}
\rho_\alpha = i\partial_\alpha \chi +f_\alpha  \ \ , \ \
\overline{\partial}_{\overline{\beta}}f_{\alpha} = 0    \label{rhogen}
\end{equation}
for some holomorphic $f_\alpha$. The symmetry~(\ref{symm}) turns out
to be the analog of the generalised K\"{a}hler gauge transformation
discussed in~\cite{GHR}. It leaves the metric and torsion invariant, but
changes
$b_{ij}$ by an
antisymmetric tensor gauge transformation of the form
$\delta b_{ij} =\partial_{[i} \lambda_{j]}$.

\section{Rigid Symmetries}   \label{kinematics}

        We now consider the isometry symmetries of the target geometry. Let
$G$ be a continuous subgroup of the diffeomorphism group of $M$. The
action of $G$ on $M$ is generated by vector fields $\xi^i_a$ ($a=1,\ldots ,
dimG$)
which satisfy the Lie bracket algebra
\begin{eqnarray}
[ \xi_a , \xi_b ]^i & \equiv & \xi^j_a \partial_j \xi^i_b - \xi^j_b \partial_j
\xi^i_a   \equiv {\cal L}_a \xi^i_b \nonumber \\
& = &  f^{c}_{ab} \xi^i_c  ,
\end{eqnarray}
where ${\cal L}_a$ denotes the Lie derivative with respect to $\xi_a$ and
$f^{a}_{bc}$ are the structure constants of the group $G$. The infinitesimal
transformations of the (1,1) sigma-model superfields
\begin{equation}
\delta \phi^i =\lambda^a \xi^i_a (\phi )        \label{rigidtransf}
\end{equation}
with constant parameters $\lambda^a$ induce a change in the (1,1)
supersymmetric
action~(\ref{S11}) which can be cancelled by the following compensating
transformations of the metric and torsion:
\begin{equation}
\delta g_{ij}=\lambda^a ({\cal L}_a g)_{ij} \ \ , \ \ \delta H_{ijk}=\lambda^a
({\cal L}_a H )_{ijk}  .     \label{dgdT}
\end{equation}
Although this is not a conventional Noether symmetry in general
(as the infinite number of coupling constants encoded in $g_{ij}$ and $H_{ijk}$
transform
under~(\ref{rigidtransf})), the action of $G$ generates a group of proper
symmetries of the sigma-model field equations if~(\ref{dgdT}) vanishes,
i.~e.\ if the Lie derivatives with respect to the vector fields $\xi^i_a$
of the metric and torsion vanish,
\begin{equation}
({\cal L}_a g )_{ij}=0 \ \ , \ \ ({\cal L}_a H)_{ijk} =0  .
\label{dgdT=0}
\end{equation}
This requires that the $\xi^i_a$ are Killing vectors of the metric $g$,
\begin{equation}
\nabla_{(i}\xi_{j)a} = 0  ,    \label{Kill}
\end{equation}
so that
$G$ is a group of isometries of $M$, and that $\xi^i_a H_{ijk}$ is curl-free,
so that there is a locally defined one-form $u_a$ such that~\cite{C&B}
\begin{equation}
\xi^i_a H_{ijk}=\partial_{[j}u_{k]a}  .          \label{dxiT=0}
\end{equation}
For the transformations~(\ref{rigidtransf}) to define a symmetry of the
sigma-model action, it is necessary in addition for $u$ to be globally
defined. The one-forms $u_a$ are only defined up to the addition of an exact
piece:
\begin{equation}
u_{ia} \rightarrow u_{ia}+\partial_i \alpha_a  .     \label{uambig}
\end{equation}
Taking the Lie derivative of~(\ref{dxiT=0}), we find that
\begin{equation}
D_{iba} = {\cal L}_b u_{ai} -f^{c}_{ba}u_{ic}    \label{defD}
\end{equation}
is a closed one-form. If it is exact, it is often possible to use the
ambiguity~(\ref{uambig}) in the
definition of $u$ to choose it to be equivariant, i.~e.\ to choose it so that
it transforms covariantly,
\begin{equation}
{\cal L}_a u_{bi}=f^{c}_{ab}u_{ic}  .        \label{uequivariant}
\end{equation}
However, in general there can be obstructions to choosing an equivariant $u$
which have an interpretation in terms of equivariant cohomology~\cite{C&B,OS}.
It will be useful to define~\cite{C&B,gauging}
\begin{equation}
c_{ab} = \xi^i_a u_{bi} .
\label{defcab}
\end{equation}

        While the conditions~(\ref{dgdT=0}) are sufficient for the isometry
generated by the Killing vector $\xi_a$ to be a symmetry of the (1,1)
supersymmetric model, a further condition is necessary in order for it to be
compatible with (2,1) supersymmetry. Under the infinitesimal transformations of
the (2,1) sigma model superfields
\begin{equation}
\delta \varphi^i = \lambda^a \xi_{a}^{i}(\varphi )  , \label{iso21}
\end{equation}
one finds that the complex structure undergoes the compensating
transformation~\cite{HullMod}
\begin{equation}
\delta J^i{}_j = \lambda^a ({\cal L}_a J)^i{}_j  .
\end{equation}
It follows that the necessary conditions for the isometry to constitute a
proper
Noether symmetry is that the metric, torsion and complex structure are
invariant
under the diffeomorphisms generated by $\xi_a$, i.~e.\ that~(\ref{dgdT=0}) and
\begin{equation}
({\cal L}_a J)^i{}_j = 0          \label{dJ=0}
\end{equation}
are satisfied. Then the $\xi_a$ are Killing vectors which are holomorphic
with respect to $J$, so that
\begin{equation}
\partial_\alpha \overline{\xi}^{\overline{\beta}}_a = 0  . \label{xiholo}
\end{equation}

        If the torsion vanishes, then $M$ is K\"{a}hler and for every
holomorphic
Killing vector $\xi^i_a$, the one-form with components $J_{ij}
\xi^j_a$ is closed so that locally there are functions $X_a$ such that $J_{ij}
\xi^j_a =\partial_i X_a$; these are the Killing potentials
which play a central role in the gauging of the supersymmetric
sigma-models without
torsion~\cite{BaggWitt,in&out}. In complex coordinates, this becomes
$\xi_\alpha =
-\partial_\alpha X_a$. When the torsion does not vanish, this generalises
straightforwardly~\cite{gauging}: if $\xi^i_a$ is a holomorphic Killing vector
field satisfying~(\ref{dxiT=0}) and~(\ref{dJ=0}), then the one-form with
components $\omega_i \equiv J_{ij}(\xi^j_a +u^j_a)$ satisfies
$\partial_{[\alpha}w_{\beta ]}=0$, so that
there are generalised complex Killing potentials $Z_a \equiv Y_a +iX_a$ such
that
\begin{equation}
\xi_{\alpha a} + u_{\alpha a} = \partial_\alpha Y_a +i\partial_\alpha X_a  .
\label{defX}
\end{equation}
The $X_a$ and $Y_a$ are locally defined functions on $M$ and they are
determined
up to the addition of constants. Their role in the construction of gauge
invariant actions for the (2,1) model will become apparent in the following
sections. Note that the transformation $u_{\alpha a} \rightarrow u_{\alpha a}
+\partial_\alpha \alpha_a$ leaves~(\ref{defX}) invariant provided $Y_a$ also
transforms as $Y_a \rightarrow Y_a +\alpha_a$. It will be useful to absorb
$Y$ into $u$, defining
\begin{equation}
u'_{\alpha a} = u_{\alpha a} -\partial_\alpha Y_a  ,
\label{defu'}
\end{equation}
so that $\xi_{\alpha a}+u'_{\alpha a} = i\partial_\alpha X_a$,
as in~\cite{gauging}; we henceforth drop the prime on $u$.

        Under the rigid symmetries~(\ref{iso21}), the variation of the
Lagrangian in~(\ref{21action}) is
\begin{equation}
\delta L = i\lambda^a \left( {\cal L}_a k_\alpha D_- \varphi^\alpha
-{\cal L}_a \overline{k}_{\overline{\alpha}}D_-
\overline{\varphi}^{\overline{\alpha}} \right)  ,
\end{equation}
where the Lie derivative of $k_\alpha$ is
\begin{equation}
{\cal L}_a k_\alpha = \xi^{\beta}_a \partial_\beta k_\alpha +
\overline{\xi}^{\overline{\beta}}_a \partial_{\overline{\beta}}k_\alpha
+k_\beta \partial_\alpha \xi^{\beta}_a  .    \label{Liek}
\end{equation}
In general the symmetries~(\ref{iso21}) will not leave the
action~(\ref{21action})
invariant; they will leave it invariant only up to a gauge transformation of
the
form~(\ref{symm}). This requires that
\begin{equation}
{\cal L}_a k_\alpha = i\partial_\alpha \chi_a +\mach_{a\alpha}
\label{ichitheta}
\end{equation}
for some real functions $\chi_a$ and holomorphic one-forms
$\mach_{a\alpha}$,
$\partial_{\overline{\beta}}\mach_{a\alpha}=0$.

        In ref.~\cite{AH1}, explicit forms for $\chi$ and $\mach$ were found
using the geometric relations reviewed in the foregoing. Several other results
concerning the relation of the isometry subgroup $G$ of $M$ to its geometry
were
also obtained. We shall now summarise these results, but the
reader is referred to~\cite{AH1} for the derivations.

        The explicit forms for $\chi$ and $\mach$ are found from~(\ref{Kill}),
(\ref{dxiT=0}), (\ref{xiholo}), (\ref{defX}),  and~(\ref{geometry}) to be
\begin{equation}
\chi_a = X_a +i\left( \overline{\xi}^{\overline{\beta}}_a
\overline{k}_{\overline{\beta}}
-\xi^{\beta}_a k_\beta \right)
\label{chi}
\end{equation}
and
\begin{equation}
\mach_{a\alpha} = 2\xi^{\gamma}_a \partial_{[\gamma}k_{\alpha ]}
+\xi_{\alpha a} -i\partial_\alpha X_a  .
\label{theta}
\end{equation}
Substituting~(\ref{chi}) and~(\ref{theta}) into~(\ref{ichitheta}), one finds
the Lie derivative~(\ref{Liek}). The holomorphy of $\mach_{a\alpha}$
in~(\ref{theta}) can be checked using~(\ref{xiholo}) and~(\ref{defX}). Note
that
the ambiguity $X_a \rightarrow X_a + C_a $ in the definition of $X_a$ (for some
constant function $C_a$) does not affect $\mach$. On the other hand, $\mach$
should transform as $\delta \mach_{\alpha a} = {\cal L}_a f_\alpha$ under
the transformations~(\ref{symm}), (\ref{rhogen}) and the form~(\ref{theta})
indeed transforms in this way.

        We also find the following expression for $u_{\alpha a}$:
\begin{equation}
u_{\alpha a} = 2\xi^{\gamma}_a \partial_{[\gamma}k_{\alpha ]}
-  \mach_{\alpha a}  .
\label{u}
\end{equation}
and this satisfies~(\ref{dxiT=0}).

        The one-form $D_{iab}$ defined by~(\ref{defD}) is closed, implying
the local existence of a real potential $E_{ba}$ such that
\begin{equation}
D_{ba\alpha} = i\partial_\alpha E_{ba}  \label{defE}
\end{equation}
(note that~(\ref{defE}) does not imply that $D_{ba}$ is exact). The potential
$E_{ba}$ is only defined up to the addition of real constants, and is
determined by the imaginary part of the generalised Killing potential. This is
seen by taking the Lie derivative of eq.~(\ref{defX}) and integrating, which
yields
\begin{equation}
E_{ba} = {\cal L}_b X_a -f^{c}_{ba} X_c +e_{ba}
\label{solE}
\end{equation}
where the $e_{ba}$ are real constants which we henceforth absorb into the
definition of $E_{ba}$.

        Using the relations~(\ref{geometry}), (\ref{xiholo}) and~(\ref{theta}),
the Lie derivative~(\ref{Liek}) of $k_\alpha$ can be written as
in~(\ref{ichitheta}) where $\mach$ and $\chi$ are given by the
forms~(\ref{theta}) and~(\ref{chi}). Further information into the relation of
the isometry subgroup $G$ of $M$ to its geometry subgroup was obtained
in ref.~\cite{AH1} by deriving the action of the Lie bracket algebra on
$k_\alpha$,
\begin{equation}
[ {\cal L}_a ,{\cal L}_b ] k_\alpha = f^{c}_{ab} {\cal L}_c k_\alpha  .
\label{algk}
\end{equation}

        First, the Lie derivatives of the potentials $\chi$ and $\mach$ satisfy
\begin{equation}
{\cal L}_b \chi_a -{\cal L}_a \chi_b - f^{c}_{ba}\chi_c = {\cal L}_b X_a
-f^{c}_{ba} X_c +i\left(
\overline{\xi}^{\overline{\beta}}_a \overline{\mach}_{\overline{\beta}b}-
\xi^{\beta}_a
\mach_{\beta b} \right)
\label{algchi}
\end{equation}
and
\begin{equation}
{\cal L}_b \mach_{\alpha a} -{\cal L}_a \mach_{\alpha b} = f^{c}_{ba}
\mach_{\alpha c} -\partial_\alpha \left( \xi^{\gamma}_a \mach_{\gamma b}
+iE_{ba} \right) .
\label{algtheta}
\end{equation}
Moreover,  eq.~(\ref{ichitheta}) implies the relation
\begin{eqnarray}
[ {\cal L}_b ,{\cal L}_a ] k_\alpha & = & f^{c}_{ba} {\cal L}_c k_\alpha
+i\partial_\alpha \left( {\cal L}_b \chi_a -{\cal L}_a \chi_b -f^{c}_{ba}\chi_c
\right) \nonumber \\
& & +\left( {\cal L}_b \mach_{\alpha a} -{\cal L}_a \mach_{\alpha b}
-f^{c}_{ba} \mach_{\alpha c} \right)
\label{actG}
\end{eqnarray}
Then it was seen that the algebras~(\ref{algchi}) and~(\ref{algtheta}),
together with eqs.~(\ref{defE}) and (\ref{solE}), imply
that the sum of the last two terms on the right hand side of~(\ref{actG})
explicitly cancels, so that~(\ref{actG}) indeed reduces to~(\ref{algk}).

        Another important consequence of~(\ref{algtheta}) follows from
symmetrization with respect to group indices: this implies that the
quantities
\begin{equation}
\hat{d}_{(ab)} \equiv \xi^{\gamma}_{(a}\mach_{\gamma b)} +iE_{(ba)}
\label{hatd}
\end{equation}
are antiholomorphic functions~\footnote{This definition corrects a sign
mistake in eq.~(54) of ref.~\cite{AH1}.},
$\hat{d}_{(ab)}=\hat{d}_{(ab)}(\overline{z})$.
Then,
defining $-c_{(ab)}$ as twice the real part of $\hat{d}_{(ab)}$, one finds
\begin{equation}
-c_{(ab)} \equiv \hat{d}_{(ab)}+\overline{\hat{d}}_{(ab)}= \xi^{i}_{(a}
\mach_{ib)} .
\label{dab}
\end{equation}
Contracting~(\ref{theta}) with $\xi^{\alpha}_{a}$ (noting~(\ref{defu'}))
and using the relation~(\ref{u}),
we find
\begin{equation}
\xi^{\alpha}_{a} \mach_{\alpha b} = 2\xi^{\alpha}_{a}\xi^{\beta}_{b}
\partial_{[\beta} k_{\alpha	]} -\xi^{\alpha}_{a}u_{\alpha b} .
\label{dwE0}
\end{equation}
Then, symmetrization with respect to group indices yields
\begin{equation}
\xi^{\alpha}_{(a}\theta_{\alpha b)} = -\xi^{\alpha}_{(a}u_{|\alpha | b)}
\label{dabxiu}
\end{equation}
so that~(\ref{dab}) can be rewritten as
\begin{equation}
c_{(ab)} = \xi^i_{(a}u_{ib)} ,
\end{equation}
which is precisely the definition of the real constants $c_{(ab)}$ given
in~\cite{C&B,gauging},
where it was shown that their vanishing is a necessary condition for the
gauging of the sigma model to be possible.

        The equivariance condition on the imaginary part of the generalised
Killing potential,
\begin{equation}
{\cal L}_b X_a = f^{c}_{ba} X_c  ,      \label{Xequiv}
\end{equation}
was found in~\cite{C&B,gauging} to be another necessary condition for the
gauging
of the isometries generated by the $\xi^i_a$ to be possible. If~(\ref{Xequiv})
holds, then it follows from~(\ref{solE}) that the potential $E_{ba}$ defined
in~(\ref{defE}) is a constant and can be chosen to vanish,
\begin{equation}
E_{ba}=0  ,
\label{E=0}
\end{equation}
and that eqns.~(\ref{defD}), (\ref{algchi}) and~(\ref{algtheta}) simplify. The
equations~(\ref{defD}), (\ref{defE}) then imply that $u$ is equivariant.

        Summarizing, the action of a group  $G$ generated by the vector fields
$\xi^i_a$ as in~(\ref{rigidtransf}) is a symmetry provided the $\xi^i_a$ are
holomorphic
Killing vectors, i.~e.\ eqs.~(\ref{xiholo}) and~(\ref{Kill}) hold, so that
the metric and complex structure are   invariant, and in addition the
torsion is invariant, i.~e.\
eqs.~(\ref{dgdT=0}) and~(\ref{dJ=0}) hold. In general, the isometry symmetries
will not leave the potential $k_\alpha$ invariant, but will change it by
a gauge transformation of the form~(\ref{symm}),
so that the action~(\ref{21action}) is unchanged.
 The geometry
and Killing potentials then determine the quantity ${\cal L}_ak_{\alpha  }$
appearing in the gauge transformation to take the form~(\ref{ichitheta})
with $\chi$, $\mach$ as in~(\ref{chi}) and~(\ref{theta}). The potentials
$\chi$ and $\mach$ satisfy~(\ref{algchi})
and~(\ref{algtheta}). Using the latter, it is  found
that the action of the
Lie bracket algebra on the vector potential $k_\alpha$
reduces to~(\ref{algk}), as  it must. Also, the quantities
$c_{(ab)}$ defined in~(\ref{dab}) are real constants equal to those
defined in ref.~\cite{gauging}. When the imaginary part of the generalised
Killing potential is chosen to be
equivariant, i.~e.\ when~(\ref{Xequiv}) holds, it is found that the potential
$E_{ba}$ defined in~(\ref{defE}) vanishes. Then the one-forms $u_a$
defined in~(\ref{defu'}) are equivariant and the geometry simplifies. We note
the result of ref.~\cite{gauging}, where it was shown that the equivariance
condition~(\ref{Xequiv}) on the imaginary part of the generalised Killing
potential must hold
in order for the gauging of the supersymmetric sigma model to be possible.

        The discussion given here also applies to the geometry and isometries
of the target space of (2,0) heterotic strings. The corresponding formulae can
be obtained from those given in the foregoing by appropriate truncation of the
(2,1) superfields.

\section{Gauging the Isometries}     \label{11gauging}

        For any supersymmetric sigma-model with a rigid isometry of the type
discussed in the previous section, one can attempt to promote the rigid
symmetry
to a local one by coupling to the appropriate super Yang-Mills multiplet. As a
warm-up,  in this section we shall briefly review the gauging of the (1,1)
model and
that of the (2,1) model in the HP formalism as presented in~\cite{gauging}.
In the following sections we
consider the gauging of the isometries of the (2,1) model in  conventional
(2,1) superspace
  formalism (adapted from~\cite{in&out}) in which extended supersymmetry
is manifest. The new   gauged actions
will be given in section~\ref{trick}.

          Let us consider first the (1,1) sigma-model with superspace
action~(\ref{S11}) and no extra supersymmetry. The (1,1) Yang-Mills
supermultiplet ${\cal A}_{(1,1)} = (A_{\pp}^{a},A_{=}^{a},
A_{+}^{a},A_{-}^{a})$ can be used to define supercovariant
derivatives $\nabla_{\pp}$, $\nabla_{=}$, $\nabla_+$ and $\nabla_-$:
\begin{eqnarray}
\nabla_{\mu}\phi^i & = & \partial_{\mu} \phi^i +A_{\mu}^{a}\xi^i_a \ \ , \ \
\mu = ( \pp , = ) \nonumber \\ \nabla_{\pm}\phi^i & = & D_{1\pm}\phi^i
+A_{\pm}^a
\xi^i_a  ,           \label{covD}
\end{eqnarray}
and these are required to satisfy the super-commutation relations~\footnote{We
use a unified notation where the super-commutator is an ordinary commutator
except when both quantities in it are anti-commuting, in which case it is the
ordinary anti-commutator.}
\begin{eqnarray}
\left[ \nabla_{+} , \nabla_{+} \right]  & = & 2i\nabla_{\pp}  \hspace{1cm}
\left[ \nabla_{-}, \nabla_{-} \right]   =  2i\nabla_{=} \nonumber \\
\left[ \nabla_{+} , \nabla_{-} \right]  & = & W \hspace{1.5cm}
\left[ \nabla_{\pp} , \nabla_{=} \right]  =  F_{\pp =} \nonumber \\
\left[ \nabla_{-}, \nabla_{\pp} \right]  & = & -i\nabla_{+} W \hspace{.5cm}
\left[ \nabla_{+}, \nabla_{=} \right]  =  -i\nabla_{-} W
\end{eqnarray}
(with all other super-commutators vanishing). The Bianchi identities imply that
$F_{\pp =}$ can be written in terms of the unconstrained field strength $W$.

        A gauge-invariant kinetic term is obtained by minimal coupling, i.~e.\
by replacing the superspace derivatives $D_{1\pm}$ by the gauge-covariant
derivatives
$\nabla_{\pm}$. The Wess-Zumino term can be gauged if there is a globally
defined $u$ satisfying eq.~(\ref{dxiT=0}), which is equivariant
(eq.~(\ref{uequivariant})) and for which
\begin{equation}
c_{(ab)} = 0 .
\label{c(ab)=0}
\end{equation}
        If all these conditions are satisfied,
then the action for the gauged (1,1) sigma-model is given in two-dimensional
form by~\cite{C&B}
\begin{eqnarray}
S & = & \int d^2 \sigma d\theta_{1+}d\theta_{1-}\left( g_{ij}\nabla_+
\phi^i \nabla_- \phi^j +b_{ij}D_{1+}\phi^i D_{1-}\phi^j \right. \nonumber \\ &
&
\left. -A_{+}^{a}u_{ia}D_{1-}\phi^i -A_{-}^{a}u_{ia}D_{1+}\phi^i +A_{-}^{a}
A_{+}^{b}c_{[ab]} \right)  .
\label{Sg11}
\end{eqnarray}

        Now suppose the conditions under which the (1,1) model with
action~(\ref{S11}) in fact has (2,1) supersymmetry hold (cf.\ section
\ref{review}). A (2,1) supersymmetric gauge-invariant action can be
constructed as follows. Let $\varphi^i$ now be (2,1) superfields and impose
the chirality constraint
\begin{equation}
D_{2+}\varphi^i = J^i{}_j D_{1+}\varphi^j ,       \label{constraint}
\end{equation}
which is equivalent to~(\ref{chiral}). Consider the HP form of
the action for the (2,1) supersymmetric model given by~\cite{HP}
\begin{equation}
S=\int d^2 \sigma d\theta_{1+}d\theta_{1-} g_{ij} D_{1+}\varphi^i
D_{1-}\varphi^j +\int d^2 \sigma dt d\theta_{1+}d\theta_{1-}
H_{ijk}\partial_t \varphi^i D_{1+}\varphi^j D_{1-}\varphi^k
\label{SHP}
\end{equation}
where the $\varphi^i (\sigma^\mu ,\theta_+ ,\overline{\theta}_+ ,\theta_- ,t)$
are interpolating superfields satisfying
\begin{eqnarray}
\varphi^i (\sigma^\mu ,\theta_+ ,\overline{\theta}_+ ,\theta_- ,0) & = & 0
\ \ , \ \ \varphi^i (\sigma^\mu ,\theta_+ ,\overline{\theta}_+ ,\theta_- ,1)
= \varphi^i (\sigma^\mu ,\theta_+ ,\overline{\theta}_+ ,\theta_- ) \nonumber \\
D_{2+}\varphi^i (\sigma^\mu ,\theta_+ ,\overline{\theta}_+ ,\theta_- ,t) & = &
-iJ^i{}_j D_{1+}\varphi^j (\sigma^\mu ,\theta_+ ,\overline{\theta}_+ ,\theta_-
,t)
 .
\end{eqnarray}
Using the chirality constraint~(\ref{constraint}) (or~(\ref{chiral})), it is
straightforward to show that the action~(\ref{SHP}) is independent of the extra
supercoordinate $\theta_{2+}$ (i.~e.\ $\delta S / \delta \theta_{2+} =
\int d^2 \sigma d\theta_{1+}d\theta_{1-} D_{2+}L =0$ up to surface
terms) provided eq.~(\ref{Jcovconst}) holds, which implies that it is
invariant under the non-manifest extra supersymmetry generated by the
supercharge $Q_{2+}$.

        To construct the gauged (2,1) supersymmetric action in this formalism,
one replaces the chiral constraints~(\ref{constraint}) by the gauge covariant
constraints
\begin{equation}
\nabla_{2+} \varphi^i = J^i{}_j \nabla_{1+} \varphi^j
\end{equation}
where the gauge covariant derivatives are defined using the
super-connections of the (2,1) Yang-Mills supermultiplet,
 which will be described in the next section. Then, using the
defining properties of the (2,1) Yang-Mills
supermultiplet and super-curvatures $W$ and $\overline{W}$, it can be shown
that
the action~\cite{gauging}
\begin{eqnarray}
S & = & \int d^2 \sigma d\theta_{1+}d\theta_{1-}\left[ g_{ij}
\nabla_{1+}\phi^i \nabla_{1-}\phi^j -i\frac{1}{\sqrt{2}} X_a \left( W^a
-\overline{W}^a \right) \right] \nonumber \\ & & +\int d^2 \sigma
dt d\theta_{1+} d\theta_{1-} \left[ H_{ijk}\nabla_t \phi^i
\nabla_{1+}\phi^j \nabla_{1-} \phi^k -\frac{1}{\sqrt{2}}
u_{ia}\nabla_t \phi^i \left( W^a +\overline{W}^a \right) \right]
\label{SgHPS}
\end{eqnarray}
is (2,1) supersymmetric provided that the conditions~(\ref{ghermJ})
and~(\ref{defX}) are satisfied. This action is also
gauge-invariant provided that $X$ and $u$ are equivariant,
i.~e.\ if~(\ref{uequivariant}) and~(\ref{Xequiv}) hold. Further, if
the conditions~(\ref{uequivariant}) and~(\ref{c(ab)=0}) hold, then the
field equations are two-dimensional and the action~(\ref{SgHPS}) can
be put in a two-dimensional gauge-invariant form similar to~(\ref{Sg11}).

\section{The (2,1) Gauge Multiplet and Gauge Symmetries}     \label{SYMetc}

        We now wish to discuss the gauging of the isometries in a formalism
which is manifestly (2,1) supersymmetric. The aim then is to promote the local
isometry symmetries~(\ref{iso21}) to local ones in which the
constant parameters $\lambda^a$ are replaced by (2,1) superfields $\Lambda^a$,
\begin{equation}
\delta \varphi^\alpha = \Lambda^a \xi^{\alpha}_a \ \ , \ \ \delta
\overline{\varphi}^{\overline{\alpha}} = \overline{\Lambda}^a
\overline{\xi}^{\overline{\alpha}}_{a} .
\label{iso}
\end{equation}
In order to ensure that these transformations preserve the chirality
constraints~(\ref{chiral}), one must require the $\Lambda^a$ to be chiral
superfields,
\begin{equation}
\overline{D}_{+}\Lambda^a = 0 \ \ , \ \ D_+ \overline{\Lambda}^a = 0  .
\end{equation}
Under a finite transformation,
\begin{equation}
\varphi \rightarrow \varphi ' = e^{L_{\Lambda \cdot \xi}}\varphi \ \ , \ \
\overline{\varphi} \rightarrow \overline{\varphi} '
= e^{L_{\overline{\Lambda} \cdot \overline{\xi}}}\overline{\varphi}  ,
\label{isoloc}
\end{equation}
where
\begin{equation}
\Lambda \cdot \xi \equiv \Lambda^a \xi^{\alpha}_{a}\frac{\partial}{\partial
\varphi^{\alpha}}
\end{equation}
and $L_{\Lambda \cdot \xi}\varphi^{\alpha}$ denotes the action of the
infinitesimal diffeomorphism with parameter $\Lambda \cdot \xi$,
\begin{equation}
L_{\Lambda \cdot \xi}\varphi^{\alpha} \equiv \left[ \Lambda \cdot \xi ,
\varphi \right]^{\alpha}  ,
\end{equation}
and acts on tensors as the Lie derivative with respect to $\Lambda \cdot \xi$.

        The (2,1) super Yang-Mills multiplet is given in (2,1) superspace
by a set of super-connections ${\cal
A}_{(2,1)}=(A_{1+}^{a},A_{2+}^{a},A_{-}^{a},
A_{\pp}^{a},A_{=}^{a})$, and these can be used to define gauge covariant
derivatives $\nabla_{1+}$, $\nabla_{2+}$, $\nabla_-$ and $\nabla_{\pp}$,
$\nabla_=$ as in~(\ref{covD}). It is convenient (cf.\ section~\ref{review}) to
combine $A_{1+}^{a}$ and $A_{2+}^{a}$ into a complex superconnection. We
define $A_{+}^{a}=\frac{1}{\sqrt{2}} (A_1^a +iA_2^a )$ and its complex
conjugate $\overline{A}_{+}^{a}$. Then ${\cal A}_{(2,1)}=(A_{+}^{a},
\overline{A}_{+}^{a},A_{-}^{a},A_{\pp}^{a},A_{=}^{a})$ and the corresponding
covariant
derivatives $\nabla_+$, $\overline{\nabla}_{+}$, $\nabla_-$, $\nabla_{\pp}$
and $\nabla_=$ satisfy the algebra
\begin{eqnarray}
\left[ \nabla_{+} , \nabla_{+} \right] & = & 0 \hspace{1cm} \left[
\overline{\nabla}_{+} , \overline{\nabla}_{+} \right]  =  0 \nonumber \\
\left[ \nabla_{+} , \overline{\nabla}_{+} \right] & = &  2i\nabla_{\pp}
\hspace{.5cm} \left[ \nabla_{-} , \nabla_{-} \right]  =  2i\nabla_{=}
\nonumber \\ \left[ \nabla_{+} , \nabla_{-} \right] & = & W \hspace{1cm} \left[
\overline{\nabla}_{+} ,\nabla_{-} \right]  =  \overline{W} \nonumber \\
\left[ \nabla_{\pp} ,\nabla_{=} \right] & = & F_{\pp =} .
\label{21alg}
\end{eqnarray}

        The super-curvatures on the right-hand side of these super-commutators
are not all independent, as they are constrained by the Bianchi identities. For
example, the Bianchi identity for the
covariant derivatives $\nabla_+$, $\overline{\nabla}_{+}$, $\nabla_-$ is
\begin{equation}
\overline{\nabla}_{+}W + \nabla_+ \overline{W} = 0 .
\end{equation}
The constraints~(\ref{21alg}) can be solved to give all connections in
terms of a scalar prepotential $V^a$ and the spinorial connection $A_{-}^{a}$.
In the chiral representation, the
right-handed spinorial derivatives that appear in the
algebra~(\ref{21alg}) are given by~\cite{Roc}
\begin{equation}
\overline{\nabla}_{+} =\overline{D}_{+} \ \ , \ \ \nabla_+ = e^V D_+ e^{-V}  ,
\label{chirrep}
\end{equation}
while the left-handed covariant derivative is defined by
\begin{equation}
\nabla_- \varphi^\alpha = D_- \varphi^\alpha -A_{-}^a \xi^{\alpha}_{a}  .
\label{nabla-}
\end{equation}

	The gauge
transformations of the Yang-Mills supermultiplet are
as follows. The real superfield
prepotential $V^a$ transforms as
\begin{equation}
e^V \rightarrow e^{V '}=e^\Lambda e^V e^{-\overline{\Lambda}}
\label{eV'}
\end{equation}
under a finite transformation.  The superconnection $A_{-}^{a}$ has the
infinitesimal gauge transformation
\begin{equation}
\delta A_{-}^{a}  =  D_- \Lambda^{a} +\left[ A_- ,\Lambda \right]^a .
\label{transfA}
\end{equation}
Because the parameters $\Lambda^a$ are complex superfields, this implies
that the connection $A_{-}^{a}$ is also complex, so that a reality condition
should be imposed on it.
The complex conjugate superconnection $\overline{A}_{-}^{a}
=(A_{-}^{a} )^{*} $ transforms as
\begin{equation}
\delta \overline{A}_{-}^{a} = D_- \overline{\Lambda}^{a} +\left[
\overline{A}_{-}
, \overline{\Lambda}  \right]^{a}
\label{transfAbar}
\end{equation}
and defines the complex conjugate covariant derivative
\begin{equation}
\overline{\nabla}_- \overline{\varphi}^{\overline{\alpha}} \equiv D_-
\overline{\varphi}^{\overline{\alpha}} -\overline{A}_{-}^a
\overline{\xi}^{\overline{\alpha}}_a  .    \label{nablabar-}
\end{equation}
A natural choice for the non-vanishing supercommutators involving the
covariant derivative $\overline{\nabla}_{-}$ is then
\begin{eqnarray}
\left[ \overline{\nabla}_{+} ,\overline{\nabla}_{-} \right] & = & {\cal
\overline{W}} \ \ , \ \ \left[ \nabla_{+} ,\overline{\nabla}_{-} \right] =
{\cal W} \nonumber \\ \left[ \nabla_- ,\overline{\nabla}_{-} \right] & = &
2i\nabla_{=} \ \ , \ \ \left[ \overline{\nabla}_{-} ,\overline{\nabla}_{-}
\right] = 0 ,
\end{eqnarray}
where the field strength $\cal W$ is obtained from $W$ by replacing $A_{-}^{a}$
with $\overline{A}_{-}^{a}$.
However, $e^V \overline{\nabla}_{-} e^{-V}$ transforms in the same way as
${\nabla}_{-}$, so that it is consistent
to identify them; this
generalised reality constraint
reduces the number of degrees of freedom of the complex field $A_-$ by a factor
of two, to get the correct counting.

        As explained at the end of the previous section, the scalar fields
$\varphi , \overline{\varphi}$ tranform under the local isometry
symmetries as in~(\ref{iso}). Now let us
define (following~\cite{in&out})
\begin{equation}
\tilde{\varphi} = e^{L_{V\cdot \overline{\xi}}}\overline{\varphi}  ,
\label{deffitilde}
\end{equation}
where
\begin{equation}
L_{V\cdot \overline{\xi}} = V^a \overline{\xi}^{\overline{\alpha}}_a
\frac{\partial}{\partial \overline{\varphi}^{\overline{\alpha}} }  .
\label{LV}
\end{equation}
Then the fields $\varphi ,\tilde{\varphi}$ satisfy the covariant
chiral constraints
\begin{equation}
\overline{\nabla}_+ \varphi^\alpha = 0 \ \ , \ \ \nabla_+
\tilde{\varphi}^{\overline{\alpha}} = 0  ,
\end{equation}
and transform under the isometry symmetries as
\begin{equation}
\delta \varphi^\alpha = \Lambda^a \xi^\alpha_a \ \ , \ \ \delta
\tilde{\varphi}^{\overline{\alpha}}_a =\Lambda^a
\tilde{\xi}^{\overline{\alpha}}_{a} (\tilde{\varphi}) .
\label{dfifitilde}
\end{equation}
Note that the transformation of $\tilde{\varphi}$ involved the parameter
$\Lambda$
while that for $\overline{\varphi}$ involved $\overline{\Lambda}$. The
left-handed covariant derivative of $\tilde{\varphi}$ is
\begin{equation}
\nabla_{-} \tilde{\varphi}^{\overline{\alpha}} = D_{-}
\tilde{\varphi}^{\overline{\alpha}} -A_{-}^{a}
\tilde{\xi}^{\overline{\alpha}}_{a}
(\tilde{\varphi}) .
\label{nablafitilde}
\end{equation}

\section{The Gauging in Superspace when $\mach =0$}  \label{notheta}

        We now discuss the gauged
(2,1) model based on the approach of ref.~\cite{in&out}. In this section we
will set the stage and perform the gauging for the special class of model for
which $\mach =0$ (cf.\ eq.~(\ref{ichitheta})), following~\cite{AH1}, while the
generic gauging
will be given in the following sections.

        We start by recalling that the general action~(\ref{21action}) for the
(2,1) model as well as the metric and torsion~(\ref{geometry}) are left
invariant under the gauge transformation~(\ref{symm}) with $\rho$ taking the
form~(\ref{rhogen}). The presence of this gauge invariance implies that the
isometries~(\ref{iso}) will not in general leave the action invariant, but
will leave
it invariant only up to gauge transformations of the form~(\ref{symm}). This
is analogous to the
situation in the more familiar K\"{a}hler case where the model has in fact
(2,2) supersymmetry and the K\"{a}hler potential is left invariant up to
K\"{a}hler gauge transformations~\cite{GHR}.

        Now consider the variation of the Lagrangian
\begin{equation}
L=i\left( k_\alpha D_- \varphi^\alpha -\overline{k}_{\overline{\alpha}}
D_- \overline{\varphi}^{\overline{\alpha}} \right)     \label{L21}
\end{equation}
under the infinitesimal rigid transformations~(\ref{iso21}). It is
straightforward to
check that
\begin{equation}
\delta L = i\left( {\cal L}_a k_\alpha D_-
\varphi^\alpha -{\cal L}_a \overline{k}_{\overline{\alpha}}
D_- \overline{\varphi}^{\overline{\alpha}} \right)  , \label{dL21}
\end{equation}
where the Lie derivative of $k_\alpha$ is given by the expression~(\ref{Liek}).
The gauge invariance~(\ref{rhogen}) then requires
generically that
\begin{equation}
{\cal L}_a k_\alpha = i\partial_\alpha \chi_a +\mach_{\alpha a}
\label{dk=rho}
\end{equation}
with $\chi$ a real function and $\mach_{a\alpha}$ a holomorphic one-form
which were shown in ref.~\cite{AH1} to take the explicit forms~(\ref{chi})
and~(\ref{theta}).

        Let us consider first the special class of models for which
\begin{equation}
{\cal L}_a k_\alpha = 0 ,   \label{Lk=0}
\end{equation}
so that the Lagrangian~(\ref{L21}) and hence the action~(\ref{21action}) are
invariant under the rigid transformations~(\ref{iso21}). Then the gauged sigma
models belonging to this class are obtained by minimal coupling. This coupling
is achieved by replacing $\overline{\varphi}$ with $\tilde{\varphi}$ and
replacing the supercovariant derivative $D_-$ with the gauge covariant
derivative
$\nabla_-$ defined in~(\ref{nabla-}) and~(\ref{nablafitilde}). This gives the
Lagrangian
\begin{equation}
L_0 = i\left( k_\alpha (\varphi ,\tilde{\varphi})\nabla_- \varphi^\alpha
-\tilde{k}_{\overline{\alpha}}(\varphi ,\tilde{\varphi}) \nabla_-
\tilde{\varphi}^{\overline{\alpha}} \right)  , \label{L0}
\end{equation}
This is indeed invariant under the transformations~(\ref{transfA}),
(\ref{eV'}) and~(\ref{dfifitilde}) provided (\ref{Lk=0})
holds.

        In the remaining part of the present section, we will gauge the more
general class of models for which the
holomorphic part of the Lie derivative~(\ref{dk=rho}) vanishes, i.~e.\
the conditions
\begin{equation}
\mach_{\alpha a }=0    \label{theta=0}
\end{equation}
and
\begin{equation}
{\cal L}_a k_\alpha = i\partial_\alpha \chi_a   \label{Lk=idchi}
\end{equation}
hold for these models, while the generic case $\mach \pp 0$ will be treated
in the following sections.

        When~(\ref{theta=0}) and~(\ref{Lk=idchi}) hold, the action based
on~(\ref{L0}) is no longer gauge invariant. Using
(\ref{transfA}), (\ref{dk=rho}), (\ref{theta=0}), and the
infinitesimal variation of the fields (\ref{dfifitilde}), we find
\begin{equation}
\delta L_0 =i\Lambda^a D_- \chi_a (\varphi ,\tilde{\varphi})
+\Lambda^a A_{-}^{b} \left( \partial_\alpha \chi_a \xi^{\alpha}_{b}
(\varphi ) +\partial_{\overline{\alpha}} \chi_a 
\tilde{\xi}^{\overline{\alpha}}_b \right) (\varphi ,\tilde{\varphi}) .
\label{dL0}
\end{equation}
This can be cancelled by adding the following term to $L_0$:
\begin{equation}
\hat{L}_0 =- A_-^a \chi_{a} (\varphi ,\tilde{\varphi})  .
\label{L0hat}
\end{equation}
The expression~(\ref{chi}) of the potential $\chi$ implies that the terms
multiplying the gauge
connection $A_-$ combine to yield the generalised Killing potential $X$:
\begin{eqnarray}
L_{g}^{(0)} & = & L_0 +\hat{L}_0 \nonumber \\ & = & i\left( k_\alpha D_-
\varphi^\alpha -\tilde{k}_{\overline{\alpha}}D_-
\tilde{\varphi}^{\overline{\alpha}} \right) (\varphi , \tilde{\varphi})
- A_-^a X_a  (\varphi ,\tilde{\varphi})  .
\label{Lgauge0}
\end{eqnarray}
We claim that the action based on the Lagrangian
$L_{g}^{(0)}$ in~(\ref{Lgauge0}) is the full
gauge-invariant action for the gauged (2,1) model in the special case where
$\mach =0$ provided that the generalised Killing potential $X$ transforms
covariantly under the isometries~(\ref{iso}), i.~e.\
\begin{equation}
\delta X_a =f^{c}_{ab}\Lambda^b X_c  .
\label{Xeqagain}
\end{equation}
To see this, note that the variation of the first term in~(\ref{Lgauge0}) is
given by
\begin{eqnarray}
\delta \left[ i\left( k_\alpha D_- \varphi^\alpha
-\tilde{k}_{\overline{\alpha}}
D_- \tilde{\varphi}^{\overline{\alpha}}\right) (\varphi ,\tilde{\varphi})
\right]
& = & -\Lambda^a D_- \chi_a (\varphi ,\tilde{\varphi}) -iD_- \Lambda^a
\left( \tilde{\xi}^{\overline{\alpha}}_a k_{\overline{\alpha}}-\xi^{\alpha}_a
k_\alpha \right) (\varphi ,\tilde{\varphi}) \nonumber \\ & = & -D_- \Lambda^a
X_a (\varphi ,\tilde{\varphi})  ,
\label{dL0D}
\end{eqnarray}
the manipulations being similar to those which lead to the
expression~(\ref{dL0}); the last identity follows upon integrating by parts,
discarding a surface term
and using the expression~(\ref{chi}) (notice that the
second term on the right-hand side of~(\ref{chi}) has cancelled). On the other
hand, the variation of the second term in~(\ref{Lgauge0}) is
\begin{equation}
\delta \left[ - A_{-}^a X_a  (\varphi ,\tilde{\varphi})
\right] = -D_- \Lambda^a X_a (\varphi ,\tilde{\varphi})
-A_{-}^a \left( \delta X_a +f^{c}_{ba}\Lambda^b X_c \right) (\varphi ,
\tilde{\varphi}) ,
\label{dL0hat}
\end{equation}
where we have used the variations~(\ref{transfA}) of the
gauge connection. Adding
(\ref{dL0D}) and~(\ref{dL0hat}), a cancellation occurs, and one is left with
\begin{equation}
\delta L_{g}^{(0)} = -A_{-}^{a} \left( \delta X_a + f^{c}_{ba}\Lambda^b X_c
\right)
(\varphi ,\tilde{\varphi})  .
\label{dLg0}
\end{equation}
This cannot be cancelled by the variation of any further addition to the action
or transformation rules,
but vanishes if $\delta X_a + f^{c}_{ba}\Lambda^b X_c=0$,
so we find that
the equivariance condition~(\ref{Xeqagain}) is
a necessary condition for the gauging to be possible.
 As seen in
section~\ref{kinematics}, the condition~(\ref{Xeqagain}) implies
the equivariance
of $u$.
Here the relation~(\ref{dab}) together with
the assumption that $\mach =0$
 implies $c_{(ab)}=0$. Thus the
conditions necessary for gauging to be possible
found in~\cite{C&B,gauging}
(the equivariance of $X$ and $u$, and $c_{(ab)}=0$)
are all satisfied.

        Summarizing, we find that, in the special case where $\mach =0$, the
action~(\ref{21action}) for the (2,1) model can be gauged provided the same
geometric
condition as that found in ref.~\cite{gauging} is satisfied, namely the
equivariance
of the generalized Killing potential $X$. Moreover, if~(\ref{Xeqagain}) holds,
then the gauged (2,1) sigma-model action in this case is the superspace
integral
of the gauge invariant Lagrangian~(\ref{Lgauge0}).

\section{Noether Gauging in Superspace}    \label{Noether}

        Let us now turn to the discussion of the gauging in the generic
situation where the holomorphic part of the Lie derivative~(\ref{dk=rho}) is
arbitrary, i.~e.\ $\mach$ may not vanish. We shall first use the Noether method
to obtain the
gauging to lowest and first order in
the gauge coupling constant $q$ in order to gain some insight into the
structure
of the full all-orders gauge invariant action. In the next section, we will
introduce a
procedure which  enables us to reduce the analysis to that of the
special case $\mach =0$ up to some subtleties which will be discussed in
detail.

        We now reinstate the gauge coupling constant $q$, which has until
now been set to 1, and note the following
first order infinitesimal variation:

\begin{equation}
q\delta V^a  =  \Lambda^a -\overline{\Lambda}^a +  \frac{q}{2}f^{a}_{bc}V^b
\left( \Lambda^c + \overline{\Lambda}^c \right) + O ( q^{2}) \label{dV}
\end{equation}
which is the infinitesimal form of~(\ref{eV'}).
Taylor expanding~(\ref{deffitilde})
gives
\begin{equation}
  \tilde{\varphi}^{\overline{\alpha}} = \bar \varphi^{\overline{\alpha}} +
q V^a
\overline{\xi}^{\overline{\alpha}}_a +\frac{q^{2}}{2} V^a V^b
\xi^{\overline{\beta}}_b \partial_{\overline{\beta}}
\xi^{\overline{\alpha}}_a		+O (q^{3}) .
\label{delfitilde}
\end{equation}
We shall also need the infinitesimal gauge transformations~(\ref{transfA}),
(\ref{transfAbar})
of the superconnections $A_{-}^{a}$ and $\overline{A}_{-}^{a}$. Note
however that, when gauging an abelian isometry subgroup, it is sufficient
to consider the lowest order variations of the gauge
fields since the structure constants $f^{a}_{bc}$ are zero in that case.

We start with the Lagrangian
$L_{g}^{(0)}$
of         the previous section, given in eq.~(\ref{Lgauge0}), which was the
full gauged Lagrangian when
$\mach = 0$. In the general case, its variation depends on $\mach $.
Under the variations~(\ref{transfA}) and~(\ref{dV}) of the
gauge superconnections and~(\ref{dfifitilde}) of the superfields $\varphi$ and
$\tilde{\varphi}$, we find
\begin{eqnarray}
\delta L_{g}^{(0)} & = & -i\Lambda^a \overline{\mach}_{\overline{\alpha}a}D_-
\overline{\varphi}^{\overline{\alpha}} -i q \Lambda^a D_- V^b
\xi^{\overline{\alpha}}_b
\overline{\mach}_{\overline{\alpha}a} -i q \Lambda^a V^b {\cal L}_b
\overline{\mach}_{\overline{\alpha}a}D_- \overline{\varphi}^{\overline{\alpha}}
\nonumber \\ & & - q A_{-}^a \left( \delta X_a +f^{c}_{ba} \Lambda^b X_c
\right)				+O (q^{2} )
\label{varLg0}
\end{eqnarray}
up to surface terms.
The term independent of $q$ can be cancelled by adding
\begin{equation}
\tilde{L}_1 = i q V^a \tilde{\mach}_{\overline{\alpha}a}(\tilde{\varphi})
\nabla_- \tilde{\varphi}^{\overline{\alpha}}
\label{L1}
\end{equation}
which can be expanded to second order in $q$ using the expansion of
$\tilde{\varphi}$ in~(\ref{deffitilde}), yielding
\begin{eqnarray}
\tilde{L}_{1}^{(2)} & = & i q V^a \overline{\mach}_{\overline{\alpha}a}D_-
\varphi^{\overline{\alpha}} +i q^{2} V^a D_- V^b \xi^{\overline{\alpha}}_b
\overline{\mach}_{\overline{\alpha}a} \nonumber \\ & & + i q^{2}
V^a V^b {\cal L}_{(a}
\overline{\mach}_{\overline{\alpha}b)}D_- \varphi^{\overline{\alpha}}
-i q^{2} V^a A_{-}^b \xi^{\overline{\alpha}}_b
\overline{\mach}_{\overline{\alpha}a} + O (q^{3} )
\label{L1e2}
\end{eqnarray}
where we have used the definition~(\ref{nabla-}) of the left-handed covariant
derivative.

Then  the gauge invariant Lagrangian to order $q^2$ is of the form
\begin{equation}
L_{g}^{(2)} = L_{g}^{(0)} +L_{1}^{(2)}  \label{Lge2}
\end{equation}
with $L_{g}^{(0)}$ the Lagrangian~(\ref{Lgauge0}), which is gauge invariant to
that
order when $\mach = 0$ and the equivariance condition~(\ref{Xequiv})
holds, and $L_{1}^{(2)}$ a second-order Lagrangian which includes the
term~(\ref{L1e2}).
The variation to order $q$ is now given by
\begin{eqnarray}
\delta \left( L_{g}^{(0)} +\tilde{L}_{1}^{(2)} \right) & = & -i q
\Lambda^a D_- V^b
\xi^{\overline{\alpha}}_{b} \overline{\mach}_{\overline{\alpha}a}
-i q \Lambda^a V^b {\cal L}_b \overline{\mach}_{\overline{\alpha}a}D_-
\varphi^{\overline{\alpha}} + \frac{q}{2}  f^{a}_{bc}V^b ( \Lambda^c +
\overline{\Lambda}^{c}  ) \overline{\mach}_{\overline{\alpha}a}D_-
\varphi^{\overline{\alpha}} \nonumber \\ & & + i q V^a
\partial_{\overline{\beta}}
\overline{\mach}_{\overline{\alpha}a}\overline{\Lambda}^{b}
\xi^{\overline{\beta}}_{b} D_- \varphi^{\overline{\alpha}} +i q V^a
\overline{\mach}_{\overline{\alpha}a} D_- ( \overline{\Lambda}^{b}
\xi^{\overline{\alpha}}_{b} ) \nonumber \\ & & +i q ( \Lambda^a
-\overline{\Lambda}^a ) D_- V^b \xi^{\overline{\alpha}}_b
\overline{\mach}_{\overline{\alpha}a} +i q D_- ( \Lambda^a
-\overline{\Lambda}^a ) V^b \xi^{\overline{\alpha}}_a
\overline{\mach}_{\overline{\alpha}b} \nonumber \\ & & +2i q (
\Lambda^a -\overline{\Lambda}^a ) V^b {\cal L}_{(a}
\overline{\mach}_{\overline{\alpha}b)}D_- \varphi^{\overline{\alpha}} \nonumber
\\ & & -i q ( \Lambda^a -\overline{\Lambda}^a ) A_{-}^b
\xi^{\overline{\alpha}}_b \overline{\mach}_{\overline{\alpha}a}
-i q V^a D_- \Lambda^b \xi^{\overline{\alpha}}_b
\overline{\mach}_{\overline{\alpha}a} \nonumber \\ & & - q A_{-}^b
\left( \delta X_a +f^{c}_{ba} \Lambda^b X_c \right)	+O (q^{2})
\label{varL1tilde}
\end{eqnarray}
(up to surface terms) which must be cancelled by the variation of additional
contributions to  the lagrangian (modulo certain geometric conditions stated
below). It
turns out that three such contributions are needed, namely
\begin{equation}
\hat{L}_{1}^{(2)} = -iq^{2} V^a A_{-}^b \xi^{\overline{\alpha}}_a
\overline{\mach}_{\overline{\alpha}b} -\frac{i}{2} q^{2} V^a V^b {\cal L}_{(a}
\overline{\mach}_{\overline{\alpha}b)}D_- \varphi^{\overline{\alpha}}
-\frac{i}{2} q^{2} V^a D_- V^b \xi^{\overline{\alpha}}_b
\overline{\mach}_{\overline{\alpha}a}  + O (q^{3}).
\label{L1hat}
\end{equation}
Varying as in~(\ref{transfA}), (\ref{dfifitilde}) and~(\ref{dV}), it can be
checked that the choice
\begin{eqnarray}
L_{1}^{(2)} & = & \tilde{L}_{1}^{(2)} + \hat{L}_{1}^{(2)} \nonumber \\ & = & i
q
V^a \overline{\mach}_{\overline{\alpha}a}D_- \varphi^{\overline{\alpha}}
+\frac{i}{2} q^{2} V^a D_- V^b \xi^{\overline{\alpha}}_b
\overline{\mach}_{\overline{\alpha}a}  +\frac{i}{2} q^{2} V^a V^b
{\cal L}_{(a}\overline{\mach}_{\overline{\alpha}b)}D_-
\varphi^{\overline{\alpha}} \nonumber \\ & & +2i q^{2} V^a A_{-}^b
\overline{\hat{d}}_{(ab)}	+O
(q^{3})
\label{L1gauge}
\end{eqnarray}
gives a Lagrangian $L_{g}^{(2)}$ in~(\ref{Lge2}) that is   gauge invariant  to
first order in $q$ (up to surface terms)
when
the equivariance condition~(\ref{Xequiv})  and the condition
\begin{equation}
\hat{d}_{(ab)} = \xi^{\alpha}_{(a}
\mach_{\alpha b)} = 0  .
\label{dab=0}
\end{equation}
hold. The latter condition is in fact equivalent to that of vanishing
constants $c_{(ab)}$ (eq.~(\ref{c(ab)=0})). To see this, recall the defining
equation for the Killing potential $X_a$,
\begin{equation}
J_{ij} \left( \xi^{j}_{a} +u^{j}_{a} \right) = \partial_i X_a .
\label{JxiuX}
\end{equation}
Contracting~(\ref{JxiuX}) with $\xi^{i}_{b}$, we find
\begin{equation}
-J_{ij} \xi^{i}_a \xi^j_b + J_{ij} \xi^i_b u^j_a = {\cal L}_b X_a .
\label{xiJetal}
\end{equation}
If $X_a$ is equivariant, i.~e.\ if the condition~(\ref{Xequiv}) holds, then
the right-hand side of~(\ref{xiJetal}) equals $f^{c}_{ba}X_c $ and,
symmetrizing with respect to group indices, we find the condition
\begin{equation}
J_{ij} \xi^{i}_{(b}u^{j}_{a)} = 0 ,
\end{equation}
which implies
\begin{equation}
\xi^{\alpha}_{(b}u_{| \alpha | a)}	=
\overline{\xi}^{\overline{\alpha}}_{(b}
\overline{u}_{| \overline{\alpha}| a)} .
\label{holisahol}
\end{equation}
It follows that the constants $c_{(ab)}$ are given by
\begin{equation}
c_{(ab)} = \xi^{i}_{(a}u_{i b)} = 2\xi^{\alpha}_{(a}u_{| \alpha | b)}	.
\label{cis2xiu}
\end{equation}
Then, using eqs.~(\ref{dabxiu}) and~(\ref{cis2xiu}), we find that the
quantities $\hat{d}_{(ab)}$ are related to the constants $c_{(ab)}$ as
follows
\begin{equation}
\hat{d}_{(ab)} = \xi^{\alpha}_{(a}\mach_{|\alpha |b)} =
-\xi^{\alpha}_{(a}
u_{|\alpha | b)} = -\frac{1}{2}c_{(ab)} .
\label{ds&cs}
\end{equation}
Thus the condition~(\ref{dab=0}) of vanishing $\hat{d}_{(ab)}$ is equivalent
to that of vanishing constants $c_{(ab)}$, eq.~(\ref{c(ab)=0}).

        It is important to notice that, apart from the contribution involving
the
$\overline{\hat{d}}_{(ab)}$ (which, as shown above, vanishes when the
condition~(\ref{c(ab)=0})
holds), the net effect
of adding the contribution $\hat{L}_{1}^{(2)}$ to $\tilde{L}_{1}^{(2)}$
is to modify some of the
numerical coefficients multiplying the individual terms in the latter; this
yields the specific coefficients appearing in~(\ref{L1gauge}). We shall see
shortly that this behaviour is generic and would be observed at each order in
the perturbative expansion: the necessity of substracting terms such as those
appearing in $\hat{L}_{1}^{(2)}$ from those in $\tilde{L}_1$ in order to obtain
a gauge
invariant result to the order considered reflects the general structure of the
all-orders gauged action, to the construction of which we now turn.

\section{General Gauging in Superspace}    \label{trick}

{}From the analysis of section~\ref{notheta}, we know that, given any local
one-form
$w_\alpha$
such that
\begin{equation}
{\cal L}_a w_\alpha =i\partial_\alpha W_a       \label{Liew}
\end{equation}
for some $W_a$, then the Lagrangian
\begin{equation}
L=i\left( w_\alpha \nabla_- \varphi^\alpha -\tilde{w}_{\overline{\alpha}}
\nabla_- \tilde{\varphi}^{\overline{\alpha}}\right) (\varphi ,\tilde{\varphi})
-q A_{-}^{a}{W}_a (\varphi ,\tilde{\varphi})
\label{Lw}
\end{equation}
is gauge invariant provided the \lq Killing potential'
\begin{equation}
X_a^{(w)} \equiv W_a -i\left( \overline{\xi}^{\overline{\alpha}}_a
\overline{w}_{\overline{\alpha}} -\xi^{\alpha}_a
w_\alpha \right)
\label{defW}
\end{equation}
is equivariant, i.~e.\ it satisfies
eq.~(\ref{Xequiv}).
The lagrangian  ~(\ref{Lw}) can be rewritten as
\begin{equation}
L=i\left( w_\alpha D_- \varphi^\alpha -\tilde{w}_{\overline{\alpha}}
D_- \tilde{\varphi}^{\overline{\alpha}}\right) (\varphi ,\tilde{\varphi})
-q A_{-}^{a}X_a^{(w)} (\varphi ,\tilde{\varphi}) .
\label{Lww}
\end{equation}
The potential $k_\alpha$ does not satisfy~(\ref{Liew}) as its Lie derivative is
given by~(\ref{ichitheta}). In the spirit of~\cite{in&out},
we seek a \lq correction' to $k_\alpha$ such that~(\ref{Liew}) is satisfied,
but the correction does not modify the geometry.
We therefore seek  to define
\begin{equation}
w_\alpha = k_\alpha -\kappa_\alpha      \label{wkk}
\end{equation}
which satisfies~(\ref{Liew}) for some $\kappa_\alpha$, which will be locally
defined in general.
Then the Lie derivative of $\kappa_\alpha$   is determined by
(\ref{ichitheta}), (\ref{Liew}) and~(\ref{defW}) to be of the form
\begin{equation}
{\cal L}_a \kappa_\alpha = \mach_{\alpha a} +2i\partial_\alpha \gamma_a
\label{cond2a}
\end{equation}
with
\begin{equation}
\gamma_a = \Im \left( \xi_a \cdot \kappa \right) = \frac{1}{2i} \left(
\xi^{\alpha}_a \kappa_\alpha -\overline{\xi}^{\overline{\alpha}}_a
\overline{\kappa}_{\overline{\alpha}}
\right)  .
\label{gammkapag}
\end{equation}

If in addition $\kappa_\alpha$ is holomorphic, then the replacement $k_\alpha
\to w_\alpha $ leaves the {\it ungauged}
action~(\ref{21action}) unchanged, so that $\kappa_\alpha$ does not change the
sigma-model
geometry. As we shall see below,
this holomorphy also results in
the elimination of $\kappa$ from the gauged action.
  We now show that although
the auxiliary   field
$\kappa_\alpha$   will not
in general exist globally, its local existence will be guaranteed by standard
arguments provided the geometric conditions~(\ref{Xequiv}) and~(\ref{c(ab)=0})
hold. We emphasize that our final results are  independent of
$\kappa_\alpha$.

        We shall seek a field
$\kappa_\alpha$ which
satisfies the following two conditions: (i) $\kappa_\alpha$ is holomorphic,
i.~e.\
\begin{equation}
\partial_{\overline{\beta}}\kappa_\alpha  =0      \label{cond1}
\end{equation}
and (ii) the Lie derivative of $\kappa_\alpha$ with respect to $\xi_a$ is
given by
\begin{equation}
{\cal L}_a \kappa_\alpha = \mach_{\alpha a} +2i\partial_\alpha \gamma_a
\label{cond2}
\end{equation}
for some real function $\gamma_a$. The integrability conditions on
$\kappa_\alpha$ following from eqs.~(\ref{cond1})
and~(\ref{cond2}) will now be derived.

        Taking the Lie derivative of~(\ref{cond2})
with respect to $\xi_b$ and antisymmetrizing with respect
to group indices yields
\begin{equation}
{\cal L}_{[b} {\cal L}_{a]} \kappa_\alpha = {\cal L}_{[b}\mach_{\alpha a]}
+2i\partial_\alpha \left( {\cal L}_{[b}\gamma_{a]}\right) .
\label{LLkap}
\end{equation}
The left-hand side of this equation can be rewritten using the Lie algebra
of $G$ and eq.~(\ref{cond2}), while the first term on the right-hand side
is given by~\cite{AH1}
\begin{equation}
{\cal L}_{[b}\mach_{\alpha a]} = \frac{1}{2}f^{c}_{ba}\mach_{\alpha c}
-\frac{1}{2}\partial_\alpha \left( \xi_{a}^{\gamma}\mach_{\gamma b}
\right) ,
\label{thetalg'}
\end{equation}
where we have set the potential $E_{ba}$ to zero, as in eq.~(\ref{E=0}); this
is possible provided the equivariance condition~(\ref{Xequiv}) on the imaginary
part of the Killing potential holds. Substituting~(\ref{thetalg'})
into~(\ref{LLkap}), we find upon integration of
the resulting equation that the compatibility of the condition~(\ref{cond2})
with the equivariance condition~(\ref{thetalg'}) requires the function
$\gamma_a$ to satisfy
\begin{equation}
{\cal L}_{[a}\gamma_{b]} = \frac{1}{2}f^{c}_{ab}\gamma_c -\frac{i}{4}\left(
\xi^{\gamma}_b\mach_{\gamma a} -\overline{h}_{ab} (\overline{z})\right)
\label{eqgamma}
\end{equation}
for some antiholomorphic function $\overline{h}$.

        It turns out that there is a simple solution to the
condition~(\ref{eqgamma}), namely
\begin{equation}
\gamma_a = \Im \left( \xi_a \cdot \kappa \right) = \frac{1}{2i} \left(
\xi^{\alpha}_a \kappa_\alpha -\overline{\xi}^{\overline{\alpha}}_a
\overline{\kappa}_{\overline{\alpha}}
\right)  .
\label{gammkap}
\end{equation}
Taking the Lie derivative of~(\ref{gammkap}) with respect to $\xi_b$,
substituting eq.~(\ref{cond2}) and antisymmetrizing with respect to group
indices, we find that $\gamma_a$ in~(\ref{gammkap}) solves the
condition~(\ref{eqgamma}) provided the antiholomorphic function $\overline{h}$
takes the form
\begin{equation}
\overline{h}_{ab}=\overline{\xi}^{\overline{\alpha}}_a
\overline{\mach}_{\overline{\alpha}b}					 .
\label{tilh}
\end{equation}

        In what follows we shall suppose that the vector field $\kappa_\alpha$
satisfies~(\ref{cond2}) with the function $\gamma_a$ as
in~(\ref{gammkap}). Then the
assumed holomorphy of $\kappa_\alpha$ implies the condition
\begin{equation}
{\cal L}_a \kappa_\alpha = \mach_{\alpha a} +\partial_\alpha \left(
\xi^{\beta}_a \kappa_\beta \right) .
\label{cond2'}
\end{equation}
In fact, no loss of generality in the following arguments will be
involved in replacing condition~(\ref{cond2}) with condition~(\ref{cond2'}).
This
can be seen by checking the compatibility of eq.~(\ref{cond2'}) with
eq.~(\ref{thetalg'}) as follows. Taking the Lie derivative of~(\ref{cond2'})
with
respect to $\xi^i_b$ and antisymmetrizing with respect to group indices, we
find
\begin{equation}
\left[ {\cal L}_{b} ,{\cal L}_a \right] \kappa_\alpha = 2{\cal L}_{[b}
\mach_{\alpha a]} +2\partial_\alpha \left\{ f^{c}_{ba} \xi^{\beta}_{c}
\kappa_{\beta} +\xi^{\beta}_{[a}\mach_{\beta b]} +\xi^{\beta}_{[a}
\partial_\beta \left( \xi^{\gamma}_{b]} \kappa_\gamma \right) \right\} .
\label{comp1}
\end{equation}
Using the Lie algebra of $G$, eq.~(\ref{cond2'}) and the following relation
\begin{equation}
\xi^{\alpha}_{[a} {\cal L}_{b]} \kappa_\alpha = \xi^{\alpha}_{[a}\mach_{\alpha
b]} +\xi^{\alpha}_{[a}\partial_\alpha \left( \xi^{\beta}_{b]}\kappa_\beta
\right)
\label{comp2}
\end{equation}
(which follows from~(\ref{cond2'}) upon contracting with $\xi^{\alpha}_b$ and
antisymmetrizing with respect to group indices), this can be rewritten as
\begin{eqnarray}
f^{c}_{ba}\mach_{\alpha c} -2{\cal L}_{[b}\mach_{\alpha a]} & = & f^{c}_{ba}
{\cal L}_c \kappa_\alpha -f^{c}_{ba} \mach_{\alpha c} +2\partial_\alpha
\left( \xi^{\beta}_{[a} {\cal L}_{b]} \kappa_\beta \right) \nonumber \\
& = & \partial_\alpha \left( 2\xi^{\beta}_{b}\xi^{\gamma}_{a}\partial_{[\beta}
\kappa_{\gamma ]} \right) ,
\label{comp3}
\end{eqnarray}
where the second equality follows from the definition of the Lie derivative
and the holomorphy of $\kappa_\alpha$. Moreover, we find from eq.~(\ref{comp2})
that
\begin{eqnarray}
\xi^{\alpha}_{[a}\mach_{\alpha b]} & = & \xi^{\alpha}_{[a} {\cal L}_{b]}
\kappa_\alpha -\xi^{\alpha}_{[a}\partial_\alpha \left( \xi^{\beta}_{b]}
\kappa_\beta \right) \nonumber \\ & = & 2\xi^{\alpha}_{a}\xi^{\beta}_b
\partial_{[\beta}\kappa_{\alpha ]} .
\label{comp4}
\end{eqnarray}
Substituting in eq.~(\ref{comp3}) and using the antiholomorphy of
$\overline{\hat{d}}_{(ab)}=\xi^{\overline{\alpha}}_{(a}
\overline{\mach}_{\overline{\alpha} b)}$ (cf.\ eq.~(\ref{actG}) and
below) then yields the equivariance condition~(\ref{thetalg'}).

        Now we show that the condition~(\ref{c(ab)=0}) of vanishing
constants $c_{(ab)}$ necessarily holds if a
vector field satisfying the conditions~(\ref{cond1}) and~(\ref{cond2'}) exists.
Contracting eq.~(\ref{cond2'}) with $\xi^{\alpha}_b$ and symmetrizing with
respect to group indices, we find the relation
\begin{equation}
\xi^{\alpha}_{(b}\mach_{\alpha a)} = \xi^{\alpha}_{(b} {\cal
L}_{a)}\kappa_\alpha
-\xi^{\alpha}_{(b} \partial_\alpha \left( \xi^{\gamma}_{a)}\kappa_\gamma
\right)
 .
\label{dab&kappa}
\end{equation}
Then, taking the Lie derivative as in~(\ref{Liek}) and using the assumed
holomorphy of $\kappa_\alpha$, one finds after some simple manipulations that
the right-hand side of~(\ref{dab&kappa}) vanishes identically. Hence two
integrability conditions on the vector field $\kappa_\alpha$ satisfying the
conditions~(\ref{cond1}) and~(\ref{cond2'}) are the equivariance
condition~(\ref{Xequiv}) and the condition~(\ref{dab=0}) of vanishing
$\hat{d}_{(ab)}$, which is equivalent to the condition~(\ref{c(ab)=0})
of vanishing constants $c_{(ab)}$ as shown above. Recall that~(\ref{c(ab)=0})
was found in~\cite{gauging} to be a necessary condition for the gauging of the
(2,1) supersymmetric sigma model to be possible (cf.\ section~\ref{11gauging}),
and that we required this same condition to hold in the perturbative analysis
given in the end of the previous section.

 Finally, we return to the issue of the holomorphy of $\kappa_\alpha$.
The condition~(\ref{cond2}) with $\gamma$ given by~(\ref{gammkap})
implies that $\kappa$ satisfies
\begin{equation}
\xi^{\beta}_{a} \partial_{\beta} \kappa_{\alpha} -\xi^{\beta}_{a}
 \partial_\alpha
\kappa_\beta +
\xi^{\bar \beta}_{a}( \partial_{\bar \beta} \kappa_{\alpha}+
 \partial_\alpha
\kappa_{\bar \beta})
= \vartheta_{\alpha a} ,
\label{pde}
\end{equation}
which is an inhomogeneous first-order partial differential equation for the
holomorphic vector field $\kappa_\alpha$.
Choosing adapted coordinates for one of the Killing vectors in which
$\xi^\alpha _a \partial _\alpha = \partial
/\partial z$ for some particular value of $a$, the equation~(\ref{pde})
becomes
\begin{equation}
\partial_z  \kappa_\alpha - \partial_\alpha \kappa _z +
[\partial_{\bar z} \kappa _\alpha + \partial _{\bar z} \bar \kappa _{\bar z}]
= \vartheta_{\alpha a}
\end{equation}
As $\vartheta_{\alpha a}$ is holomorphic, this can clearly be integrated for
holomorphic $\kappa$.
Thus $\kappa$ can be chosen to be holomorphic with respect to the
 coordinates corresponding to any commuting set of Killing vectors.
If the integrability conditions considered above, and in particular
  the
condition~(\ref{Xequiv}),  hold, then the equation~(\ref{pde}) will have local solutions
  $\kappa$ which are  holomorphic.
This establishes the local existence of an auxiliary vector field
$\kappa_\alpha$
with the desired properties.

        Then,
using the definitions~(\ref{defW}) and~(\ref{wkk}) and the holomorphy
of $\kappa_\alpha$, we find that the Lagrangian
{}~(\ref{Lww}), which formally is gauge invariant, can be rewritten as
\begin{eqnarray}
L & = & i\left( k_\alpha D_- \varphi^\alpha -\tilde{k}_{\overline{\alpha}}D_-
\tilde{\varphi}^{\overline{\alpha}} \right) (\varphi ,\tilde{\varphi} )
-q A_{-}^a X_a (\varphi ,\tilde{\varphi}) \nonumber \\ & & +
i\tilde{\kappa}_{\overline{\alpha}}
(\tilde{\varphi})D_- \tilde{\varphi}^{\overline{\alpha}}
\label{Lkappa}
\end{eqnarray}
where we have discarded a term $\kappa_\alpha D_- \varphi^\alpha$, which is
chiral as a result of the holomorphy of $\kappa_\alpha$. Note that  other terms
have cancelled from~(\ref{Lkappa}).

     The expression~(\ref{Lkappa}) of the gauged Lagrangian  is not
satisfactory as it involves the   vector field
$\kappa_\alpha$, which is only defined implicitly; the action obtained using
the Noether method involves no such vector. We must therefore
endeavour to rewrite it in such a way that no explicit dependence on
$\kappa$ remains. To this end, we write the last term in~(\ref{Lkappa}) in the
following way:
\begin{eqnarray}
\tilde{\kappa} (\tilde{\varphi})D_- \tilde{\varphi}^{\overline{\alpha}} & = &
e^L
\left( \overline{\kappa}_{\overline{\alpha}}(\overline{\varphi})D_-
\overline{\varphi}^{\overline{\alpha}}\right) \nonumber \\ & = &
\overline{\kappa}_{\overline{\alpha}}(\overline{\varphi})D_-
\overline{\varphi}^{\overline{\alpha}} + \frac{e^L -1}{L} L
\left[ \overline{\kappa}_{\overline{\alpha}}(\overline{\varphi})D_-
\overline{\varphi}^{\overline{\alpha}}\right]    ,
\label{ikDfi}
\end{eqnarray}
where we have defined
\begin{equation}
L \equiv {\cal L}_{V\cdot \overline{\xi}}
\label{Ltensor}
\end{equation}
and used the identity
\begin{equation}
e^L=1+\frac{e^L -1}{L} L.
\label{id}
\end{equation}
The first term in the last line of~(\ref{ikDfi}) is antichiral by holomorphy of
$\kappa_\alpha$, and can be
discarded. Moreover, a simple calculation utilizing the
definition~(\ref{Ltensor})
of the operator $L$ and the relations~(\ref{cond2}) and~(\ref{eqgamma}) yields
\begin{equation}
L \left[ \overline{\kappa}_{\overline{\alpha}}(\overline{\varphi})D_-
\overline{\varphi}^{\overline{\alpha}} \right] =  q V^a
\overline{\mach}_{\overline{\alpha}a}
D_- \overline{\varphi}^{\overline{\alpha}} +D_- \left(   q V^a
\overline{\xi}^{\overline{\alpha}}_a
\overline{\kappa}_{\overline{\alpha}} \right)  .
\end{equation}
Hence we find that the $\kappa$-dependent term in~(\ref{Lw}) can be rewritten
as
\begin{eqnarray}
i  \tilde{\kappa}_{\overline{\alpha}}(\tilde{\varphi})D_-
\tilde{\varphi}^{\overline{\alpha}} & = & i  \frac{e^L -1}{L} q V^a
\overline{\mach}_{\overline{\alpha}a}D_- \overline{\varphi}^{\overline{\alpha}}
+i   \frac{e^L -1}{L} D_- \left( q V^a \overline{\xi}^{\overline{\alpha}}_{a}
\overline{\kappa}_{\overline{\alpha}}  \right) \nonumber \\ & = & i
\frac{e^L -1}{L}      q	V^a \overline{\mach}_{\overline{\alpha}a}D_-
\overline{\varphi}^{\overline{\alpha}}
+i  D_- \left[ \frac{e^L -1}{L} q V^a \overline{\xi}^{\overline{\alpha}}_{a}
\overline{\kappa}_{\overline{\alpha}} \right]
\label{impstep}
\end{eqnarray}
where the second inequality follows from the fact that the operator $L$
in~(\ref{Ltensor})
is the generator of infinitesimal gauge transformations with parameter the
prepotential $V$, and hence must commute with the supercovariant derivative
$D_-$. As
a result, all terms in the expansion of the second term in the first line
of eq.~(\ref{impstep}) can be recast into a total derivative term, as indicated
in the second line.

        The gauge invariant action (to all orders) for the gauged (2,1)
heterotic sigma model is the superspace integral of
the Lagrangian~(\ref{Lkappa}). Hence, upon substitution of the
expression~(\ref{impstep}) in~(\ref{Lkappa}), we find the action
\begin{eqnarray}
S_g & = & \int d^2 \sigma d\theta_+ d\overline{\theta}_+ d\theta_- \left\{
\left[ i\left( k_\alpha D_- \varphi^\alpha -k_{\overline{\alpha}}D_-
\tilde{\varphi}^{\overline{\alpha}}\right) -  q A_{-}^a X_a \right] (\varphi ,
\tilde{\varphi}) \right. \nonumber \\ & & \hspace{3cm}  \left.
+i  \frac{e^L -1}{L} q V^a \overline{\mach}_{\overline{\alpha}a}D_-
\overline{\varphi}^{\overline{\alpha}} \right\} .
\label{Lfin}
\end{eqnarray}

            Several comments on this result are in order. First, it is obvious
that in the special case where $\mach=0$, ~(\ref{Lfin}) reduces to the
superspace
integral of the gauged Lagrangian~(\ref{Lgauge0}), as indeed it must. Moreover,
the geometric conditions for gauge invariance of both~(\ref{Lfin}) and
{}~(\ref{Lgauge0}) are the equivariance of the generalized Killing potential
$X_a$
and the vanishing of the constants $c_{(ab)}$ defined in~(\ref{dab}), i.~e.\
eqs.~(\ref{Xequiv}) and~(\ref{c(ab)=0}) respectively.

        Second, it can be checked using the definition~(\ref{Ltensor}) of the
operator
$L$ that the expansion of the second term in~(\ref{Lfin}) to second order
in the gauge coupling constant $q$ yields precisely those terms appearing in
the
second-order Lagrangian~(\ref{Lge2}), which was shown above (by explicit
variation) to be gauge invariant
to first order in $q$ when the geometric conditions~(\ref{Xequiv})
and~(\ref{c(ab)=0})
hold. Furthermore, we have also checked by a (somewhat lenghty) direct
calculation that the expansion
of the last term in~(\ref{Lkappa}) to second order in $q$, which contains many
terms
involving the vector field $\kappa_\alpha$ as well as its first and second
order
derivatives, can indeed be recast as the sum of the expected
terms~(\ref{L1gauge})
appearing in the second-order gauged Lagrangian~(\ref{Lge2}) and a total
derivative term, as in~(\ref{impstep}):
\begin{eqnarray}
\tilde{\kappa}_{\overline{\alpha}}(\tilde{\varphi}) D_-
\tilde{\varphi}^{\overline{\alpha}} & = & q V^a
\overline{\mach}_{\overline{\alpha}a}
D_- \overline{\varphi}^{\overline{\alpha}} +\frac{q}{2}V^a D_- V^b
\overline{\xi}^{\overline{\alpha}}_{b}\overline{\mach}_{\overline{\alpha}a}
+\frac{q^{2}}{2}
V^a V^b {\cal L}_{(b}\overline{\mach}_{\overline{\alpha}a)}D_-
\overline{\varphi}^{\overline{\alpha}} \nonumber \\
& & + D_- \left[
q V^a
\overline{\xi}^{\overline{\alpha}}_{a}\overline{\kappa}_{\overline{\alpha}}
+\frac{q^{2}}{2}V^a V^b \overline{\xi}^{\overline{\alpha}}_{(a}
\overline{\mach}_{\overline{\alpha}b)}
+\frac{q^{2}}{2}V^a V^b \overline{\xi}^{\overline{\alpha}}_{(a}
\partial_{\overline{\alpha}}
\left( \overline{\xi}^{\overline{\gamma}}_{b)}
\overline{\kappa}_{\overline{\gamma}}\right) \right] \\
& & +O(q^3)
 .
\end{eqnarray}
Using the condition~(\ref{dab=0}) and the definition~(\ref{Ltensor}), it is
easily
seen that this is precisely the expansion of eq.~(\ref{impstep}) to that order.
Notice that this derivation relies only on the two defining conditions for the
vector field $\kappa_\alpha$, namely~(\ref{cond1}) and~(\ref{cond2}) (with
$\gamma_a$ as in~(\ref{gammkap})). This is a non-trivial test
of our results, particularly of the structure given in~(\ref{impstep}).

        Summarizing, we find that the (2,1) superspace action~(\ref{21action})
can be gauged provided the geometric conditions~(\ref{Xequiv})
and~(\ref{c(ab)=0})
hold, in which case the gauged superspace action is given in~(\ref{Lfin}).
Although
our construction utilized a vector field $\kappa_\alpha$ satisfying certain
requirements, we stress that our final action~(\ref{Lfin}) is   independent
of such an object, and its gauge invariance can be checked directly.

\section{Conclusion}              \label{concl}

        The main results of this paper can be summarized as follows. If the
(2,1) sigma-model with torsion is formulated in extended superspace, as in
(\ref{S11})-(\ref{ghermJ}) and~(\ref{21action}), then its isometries can be
gauged by coupling the model to the usual (2,1) Yang-Mills supermultiplet
provided
the geometric condition~(\ref{Xequiv}) and~(\ref{c(ab)=0}) are satisfied. If
(\ref{Xequiv}) and~(\ref{c(ab)=0}) hold, then there are two alternative forms
of the gauged action: the result~(\ref{SgHPS}) was found in
ref.~\cite{gauging},
while the manifestly (2,1) supersymmetric gauged action~(\ref{Lfin}) is new.

        In the special case where the geometry of the target space is
K\"{a}hler or twisted K\"{a}hler, the (2,1) gauged action~(\ref{Lfin}) should
reduce to the (2,2) gauged actions found in~\cite{in&out,gauging}, although
we have not presented a proof here. Also, a new action for the gauged (2,0)
supersymmetric sigma model with torsion may be obtained from~(\ref{Lfin})
by appropriate truncation of the (2,1) superfields. This gives
\begin{eqnarray}
S_g & = & \int d^2 \sigma d\theta_+ d\overline{\theta}_+   \left\{
\left[ i\left( k_\alpha \partial_= \varphi^\alpha
-k_{\overline{\alpha}}\partial_=
\tilde{\varphi}^{\overline{\alpha}}\right) - q A_{=}^a X_a \right] (\varphi ,
\tilde{\varphi}) \right. \nonumber \\ & & \hspace{3cm}  \left.
+i  \frac{e^L -1}{L} q V^a \overline{\mach}_{\overline{\alpha}a}\partial_=
\overline{\varphi}^{\overline{\alpha}} \right\} .
\label{Lfin20}
\end{eqnarray}
where all fields are (2,0) superfields. Here the (2,0) superspace is
parametrised
by two bosonic null coordinates $(\sigma^{\pp} ,\sigma^{=})$ and two Grassmann
coordinates $(\theta_{1+}, \theta_{2+})$ of the same chirality; the
supercovariant derivatives $D_{1,2+} = \partial / \partial \theta_{1,2+} +
i\theta_{1,2+} \partial /\partial \sigma^{\pp }$
satisfy $D_{1,2+}^{2} =i\partial_{\pp}$ and $[ D_{1+} ,D_{2+} ] =0$.

        As was mentioned in the introduction, our work is of relevance to the
study of the geometry of the $N=(2,1)$ heterotic string theory because it shows
how the $U(1)$ current of the internal left-moving sector of that theory can be
gauged, as indeed it must be. Since the latter sector contains 8 chiral bosons,
it is important to construct a covariant action with manifest (2,1)
supersymmetry
which describes the chiral bosons off-shell as well as on-shell. This can be
achieved by coupling the (2,1) heterotic sigma model to supergravity. The
methods
of the present paper can then be applied to the gauging of the resulting
action.
This will be discussed in more detail elsewhere~\cite{prepa}.

\vspace{1.5cm}

{\bf Acknowledgements}
\\
\\
        The work of M.\ A.\ is supported by the Janggen-P\"{o}hn Stiftung,
the Board of the Swiss Federal Institutes of Technology and the Fonds National
Suisse.

\end{document}